# Cross sections of X-ray production induced on Ti, Fe, Zn, Nb and Ta by O, Cl, Cu and Br ions with energies between 4 MeV and 40 MeV.

José Emilio Prieto[1,2,*], Patricia Galán[1] and Alessandro Zucchiatti[3]

[1]*Centro de Micro Análisis de Materiales (CMAM), Universidad Autónoma de Madrid, 28049 Madrid, Spain*

[2]*Dpto. de Física de la Materia Condensada, IFIMAC and Instituto ¨Nicolás Cabrera¨, Universidad Autónoma de Madrid, 28049 Madrid, Spain*

[3]*Universidad Autónoma de Madrid, Rectorado, 28049 Madrid, Spain*

ABSTRACT

Differential cross section of X-ray production induced by O, Cl, Cu and Br ions with energies between 4 MeV and 40 MeV have been measured for thin targets of Ti, Fe, Zn, Nb and Ta in a direct way. A fully characterized silicon drift diode was used as X-rays detector. Beam currents have been measured by a system of two Faraday cups. Corrections for target thickness effects have been applied to the raw data. Experimental cross sections are compared both with theory and with previously published results. Experimental results from other authors are in reasonable agreement with ours over a wide energy range. Theory produces consistent results in the case of oxygen ions but gives cross sections even orders of magnitude below the experimental ones for heavier ions (ECPSSR-UA) or contrasting results (PWBA) depending on the ion-target combination.

# 1-INTRODUCTION

The multi-technique character of Ion Beam Analysis (IBA) allows combining Secondary Ion Mass Spectrometry (SIMS) and particle-induced X-ray emission (PIXE) performed with light and medium-mass ions. SIMS, with incident ions at MeV energies (MeV-SIMS), has been recently established as an analytical method, relatively simple and cheap to implement [1-4]. The International Atomic Energy Agency (IAEA) has promoted a Coordinated Research Project (CRP F11019) [5] meant at consolidating, MeV-SIMS as an analytical technique and developing the combined application of quantitative Heavy-Ion PIXE (HI-PIXE) analysis. With this goal, a specific effort has been made by the CRP partners to increase the availability of X-ray production cross sections [6,7], in particular for the ion-target combinations that were identified as the most interesting for a combined MeV-SIMS and HI-PIXE sample characterization. Using the measurement protocols and data analysis procedures of our previous study [7] for C and Si ions, we have extended the investigation to the production of X-rays induced by O, Cl, Cu and Br ions with energies between 4 MeV and 40 MeV on thin targets of $TiO_2$, $Fe_3O_4$, $ZnO$, $Nb_2O_5$ and $Ta_2O_5$ grown on different substrates. This corresponds to a maximum of 1.66 MeV/u in the case of O ions and a minimum of 0.12 MeV/u when Br ions are used. The X-ray production cross sections are presented and discussed.

# 2-EXPERIMENTAL SET-UP

The experimental set-up was described in connection with the measurement of cross sections for HI-PIXE induced by C and Si ions at the 5 MV tandem accelerator of the Centro de Micro-Análisis de Materiales (CMAM), Universidad Autónoma de Madrid (UAM), Spain. Full account of the beamline and end station, as well as of the detector performance and target sizes and composition can be found in a previous article [7]. We just recall that a few targets were distributed amongst the CRP partners. They have been produced by atomic layer deposition (ZnO, $TiO_2$, TiN), sputtering ($Ta_2O_5$) or reactive sputtering ($Nb_2O_5$, $RuO_2$). In addition, the targets TiN-C and $Fe_3O_4$-C were produced at UAM by ion beam sputtering on a silicon substrate, using a Kaufman-type ion source. Target thicknesses, reported in Table I, have been extracted from the analysis of elastically backscattered 2 MeV alpha particles. The estimated uncertainty of the target thicknesses is on average around 2.5%.

The charge was measured comparing the data provided by an insertion (IFC) and a transmission Faraday Cup (TFC), both with a secondary electrons suppressor. The IFC is used to determine the net charge sent to target immediately before and after the irradiation, while the TFC signal is used to record possible current fluctuations and to correct the values provided by the IFC. The TFC intercepts (7.1 ± 1.5) percent of the beam going to the sample, as determined using a 2 MeV proton beam and the ratios of the integrated charges given by the two FC. The uncertainty of the measured charge depends on the stability of current: we observed an average of 2.7% and an extreme value of 11%, considering the O, Cl, Cu and Br ions cumulatively.

X-rays are detected by a KETEK AXAS-A 10mm$^2$ silicon drift detector (SDD), at 120 degrees from the beam direction, protected by an absorber (500 microns polyethylene terephthalate PET) from the impact of backscattered ions and protons. Using a 2 MeV proton beam and the X-ray production cross sections compiled in the GUPIX software package **[8]**, we have built the efficiency curve, $\varepsilon_{abs}(E_X)$, as a function of X-ray energy. Its uncertainty is on average 4% (minimum: 2.8%; maximum: 7.2%); the difference with a third order polynomial fitting curve is on average 7%. For the K-lines we used experimental efficiencies to extract the cross sections. In the case of the Ta L-series, we used the experimental efficiency for the $L_\alpha$ line and fitted values for all other lines.

Currents between 2.5 nA and 130 nA have been used for O ions in charge states ranging from 2 to 4 with an average deadtime of 0.3% (see Table II). Currents between 4.5 nA and 110 nA have been used for Cl ions in charge states ranging from 3 to 7 with an average deadtime of 2.2%; between 0.87 nA and 2.82 nA for Cu ions in charge states 3 to 7 with an average deadtime of 2.0%; between 1.42 nA and 29.2 nA for Br ions in charge states 3 to 7 with an average deadtime of 1.1%. Preamplifier signals have been processed by an ORTEC 572A amplifier and a Fast Comtec 7072 dual analog-to-digital converter. Spectra have been collected and processed with a Fast Comtec MP3 multi-parameter system.

It is important to note that, when using Cu and Br ions, the K lines produced by excitation of the incoming ion, may mix with the target element characteristic lines. In these cases the spectra (see the example of figure 1) have been fitted by a series of Gaussian functions plus a user defined continuous background, leaving the peak centroids, widths and heights as free parameters. Doing so we have been able to account for peak energy shifts and broadening **[9]** of the target element with respect to proton PIXE and to subtract the contribution of the incoming ion peaks to the HI-PIXE spectrum. We estimated a combined standard uncertainty (including the global statistical and fitting uncertainties) on each individual yield between 1% and 5%, being the highest values observed for the less intense peaks (e.g. Nb K$_\beta$, Ta L$_{\gamma1}$) or for the peaks most affected by the superposition with the incoming ion ones (e.g. Zn K$_\alpha$ and Zn K$_\beta$ peaks excited by Cu ions).

3-DIFFERENTIAL CROSS SECTIONS

The differential cross sections for the reaction channels mentioned above are given by:

$$\frac{d\sigma_X(E_I,\vartheta)}{d\Omega} = \frac{Y_X(E_I,\vartheta)}{N_I N_T \varepsilon_{abs}(E_X) 4\pi} \cdot f_{corr} \qquad N_I = \frac{Q}{ne} \qquad N_T = \frac{N_0 \rho dx}{A} \qquad (1)$$

$Q$ is the net collected charge, $n$ is the ion charge state, $e$ is the unit charge and $N_I$ is the corresponding number of incident ions. A is the analyte atomic weight, $\rho \cdot dx$ is the analyte areal density (see Table I), $N_0$ is the Avogadro number and $N_T$ is the corresponding number of target atoms. $E_I$ is the incoming ion energy, $Y_X(E_I,\theta)$ is the X-ray yield measured at 120 degrees at that energy, while $f_{corr}$ is a correction factor due to ion energy loss and X-ray absorption in the thin targets. All quantities on the right side have been measured.

The correction factor has been calculated from equation (2). The specific energy loss [10] $S(E)$ and the theoretical ECPSSR cross sections [11-14] $d\sigma/d\Omega(E,\theta)$ have been fitted with polynomials, of at most the third order, to compute the integrals in equation (2) from the beam energy $E_I$ to the energy $E_f$ of the ion at the target exit. A complete discussion of the numerical procedure can be found elsewhere [15].

$$f_{corr} = \frac{d\sigma}{d\Omega}(E_I, \theta) \int_{E_f}^{E_I} \frac{1}{S(E)} dE \Big/ \int_{E_f}^{E_I} \frac{\frac{d\sigma}{d\Omega}(E,\vartheta)}{S(E)} dE \qquad (2)$$

The correction increases with the ion's Z number. It is of a few percent at the highest ion energies but can reach a value close to 2 for the thickest target ($Fe_3O_4$) and the lowest ion energy [15].

The experimental-set-up allowed us to measure both the $K_\alpha$ and $K_\beta$ yields in Ti, Fe, Zn at all energies. For Nb, due to the reduced detector efficiency and the lower cross sections, we were able to measure both the $K_\alpha$ and $K_\beta$ only for oxygen and the sole $K_\alpha$ in a few cases for chlorine. The results obtained for the differential cross sections are reported in Table II. In order to compare with the results of other member groups of the IAEA Coordinated Research Project and with available theories (see below), the $K_\alpha$ and $K_\beta$ values were added to give the K cumulative differential cross-section. Similarly, the Ta L-lines values were added to give the L cumulative differential cross-section.

The differential cross-sections produced by O and Cl ions are compared with theory in figures 2 and 3 respectively. It has to be observed that, while in the case of oxygen there is a good agreement among experimental data and ECPSSR theory [11-14], for the case of the United Atoms (UA) approximation there is an evident discrepancy between the experimental points and the ECPSSR-UA calculations for any other target element investigated in our work. The theory has been found satisfactory for incident alpha particles [16], beryllium [17], carbon [7, 18-20] and oxygen [21] but for increasing values of the ion's Z, it remains well below the experimental data as observed also by other authors [22,23]. We observe in figure 3 that the Plane- Wave Born Approximation (PWBA) calculation overestimates the experimental cross section for any target in the case of Cl ions.

Figure 2 shows furthermore that there is a reasonable agreement between our data and the few other available in literature. For the case of oxygen ions, the data of Scafes et al. [21] on Ti lie close to ours around 20 MeV but, together with those of Gorlachev et al. [24] lie lower below 20 MeV. The data of Gorlachev et al. for Zn and Nb agree reasonably well with ours. In all cases our data are the ones which better agree with the results of ECPSSR-UA calculations.

For the case of chlorine ions on Ti, the data of Tanis et al. [23] lie very close to ours in the range 20 MeV - 40 MeV (they practically coincide at 40 MeV) and then continue increasing smoothly up to 60 MeV (see Figure 3). We were unable to find analogous data for Fe, Zn and Ta excited by Cl. However, if we consider the published cross sections in Mn (Z=25) **[23]**, we observe that they reasonably agree with our Fe (Z=26) data in the range 20 MeV - 40 MeV and again smoothly grow up to 60 MeV.

The cross sections for Cu and Br incident ions (figure 4) grow monotonically with energy and the ECPSSR-UA theoretical values (not shown) **[11-14]** lie again well below the experimental values. Both for Cu and Br incident ions, the PWBA calculation underestimates the Fe, Zn and Ta results. At the same time, PWBA overestimates the Ti results partly (for Cu ions) or totally (for Br ions).

4-CONCLUSIONS

We have applied at CMAM a measuring protocol meant at extracting differential X-ray production cross sections excited by heavy ions from fundamental parameters and a data analysis procedure that takes into account target thickness effects. The protocol, which proved its validity for the cases of C and Si ions, as reported in a previous article **[7]**, has been extended to O, Cl, Cu and Br ions, with energies of 1.66 MeV/u and below, on the same targets. Investigated elements included Ti, Fe, Zn, Nb and Ta. The differential cross section has been determined for 86 new ion-target-energy combinations. The results demonstrate first of all that the use of HI-PIXE in materials characterization is possible since it is supported by high cross sections and relatively clean spectra. However a quantitative analysis must be limited, for the moment, to selected cases where the production cross section is known experimentally and confirmed by different authors. Indeed we are still far from being able of applying an analysis protocol that tabulates cross sections (theoretical or analytical **[25]**) and fluorescence yields for any possible analyte, like it is done routinely in the case of protons and alpha particles. The theoretical ECPSSR-UA predictions, for the targets explored within IAEA's CRP, are satisfactory for oxygen ions, as well as for carbon and silicon ions. Our experimental values for oxygen are comparable to previously published ones and lie closer than others to the theoretical calculations. For increasing Z values of the incoming ion the agreement between experiments remains reasonable, but the ECPSSR-UA theory predictions lie below the experimental data with a clear difference in the case of chlorine and by orders of magnitude for the cases of Cu and Br ions. PWBA calculations give contrasting results: for Cl incident ions, they overestimate the data. Both for Cu and Br incident ions they underestimate the Fe, Zn and Ta results but partly or totally overestimate the Ti results. The inclusion of effects beyond those considered by ECPSSR (e.g. multiple ionization or electron capture) can bring some level of improvement **[9]**. However, there is increasing evidence **[26]** that different theories for calculation of X-ray production cross sections by heavy ions (PWBA, ECPSSR, Unitary Convolution Approximation, Coupled Channels, etc. give results that can differ by orders of magnitude in some cases **[26]**. These findings, together with our experimental cross sections suggest a revision of the physics of ion-target interaction for heavier ions and higher energies.

ACKNOWLEDGMENTS

Thanks are due to the CMAM technical staff for their support during the data collection and data analysis stages. We acknowledge the support received by IAEA coordinated research project (CRP) #F11019: "Development of molecular concentration mapping techniques using MeV focused ion beams" and by Project No. MAT2014-52477-C5-5-P of the Spanish MINECO. We thank all the CRP colleagues for fruitful informative discussions. Thanks are due to Pilar Prieto for the provision of targets.

TABLES

| Targets | Analyte Thickness [μg/cm$^2$] | | | Target Thickness [μg/cm$^2$] | Density [g/cm$^3$] |
|---|---|---|---|---|---|
| TiN/Si | 11.2 | ± | 0.4 | 14.5 | 4.82 |
| TiN-C/Si | 19.1 | ± | 0.5 | 28 | 4.82 |
| TiO2/Si | 11.6 | ± | 0.2 | 19.4 | 4.23 |
| ZnO/Si | 22.9 | ± | 0.4 | 29 | 5.61 |
| Fe3O4-C/Si | 64 | ± | 2 | 86 | 5 |
| Nb2O5/sigradure | 14.3 | ± | 0.2 | 21 | 4.6 |
| Ta2O5/Sigradure | 15.9 | ± | 0.4 | 19.8 | 8.18 |

**Table I.** Thickness of the used targets as determined by 2 MeV alpha backscattering. Targets TiN-C and Fe$_3$O$_4$-C were produced at UAM by ion beam sputtering.

| Energy | Charge state | Zn K [b/sr] | | | Ti K [b/sr] | | | Fe K [b/sr] | | | Ta L [b/sr] | | | Nb K [b/sr] | | |
|---|---|---|---|---|---|---|---|---|---|---|---|---|---|---|---|---|
| O ion | | | | | | | | | | | | | | | | |
| 20 | 4 | 9.58E+00 | ± | 4.83E-01 | 2.23E+02 | ± | 1.18E+01 | 5.49E+01 | ± | 3.47E+00 | 6.43E+01 | ± | 4.09E+00 | 7.37E-01 | ± | 4.02E-02 |
| 16 | 4 | 3.74E+00 | ± | 1.49E-01 | 7.06E+01 | ± | 3.66E+00 | 1.76E+01 | ± | 1.11E+00 | 2.82E+01 | ± | 1.67E+00 | 2.48E-01 | ± | 1.32E-02 |
| 12 | 4 | 1.24E+00 | ± | 4.93E-02 | 1.57E+01 | ± | 8.13E-01 | 4.33E+00 | ± | 2.73E-01 | 1.51E+01 | ± | 8.95E-01 | 8.58E-02 | ± | 4.50E-03 |
| 8 | 3 | 2.51E-01 | ± | 9.83E-03 | 2.85E+00 | ± | 1.50E-01 | 8.02E-01 | ± | 5.06E-02 | 4.48E+00 | ± | 2.61E-01 | 1.53E-02 | ± | 9.09E-04 |
| 4 | 2 | 1.13E-02 | ± | 5.03E-04 | 1.10E-01 | ± | 5.79E-03 | 2.88E-02 | ± | 1.79E-03 | 1.60E-01 | ± | 9.66E-03 | | | |
| Cl ion | | | | | | | | | | | | | | | | |
| Energy [MeV] | Charge state | Zn K [b/sr] | | | Ti K [b/sr] | | | Fe K [b/sr] | | | Ta L [b/sr] | | | Nb Ka [b/sr] | | |
| 40 | 7 | 9.41E+00 | ± | 4.41E-01 | 1.06E+03 | ± | 6.42E+01 | 1.40E+02 | ± | 9.65E+00 | 1.06E+02 | ± | 6.52E+00 | 1.69E-01 | ± | 9.93E-03 |
| 33 | 6 | 6.36E+00 | ± | 2.67E-01 | 7.96E+02 | ± | 4.42E+01 | 9.74E+01 | ± | 7.11E+00 | 7.01E+01 | ± | 5.16E+00 | 1.18E-01 | ± | 8.68E-03 |
| 26 | 5 | 1.74E+00 | ± | 6.87E-02 | 3.78E+02 | ± | 2.12E+01 | 2.79E+01 | ± | 1.99E+00 | 3.25E+01 | ± | 1.88E+00 | | | |
| 19 | 4 | 4.07E-01 | ± | 1.62E-02 | 1.46E+02 | ± | 7.29E+00 | 7.01E+00 | ± | 5.33E-01 | 1.00E+01 | ± | 5.82E-01 | | | |
| 12 | 3 | 3.47E-02 | ± | 2.66E-03 | 3.97E+01 | ± | 2.00E+00 | 8.94E-01 | ± | 7.17E-02 | 1.20E+00 | ± | 7.15E-02 | | | |
| 9 | 3 | 3.39E-03 | ± | 2.36E-04 | 1.57E+01 | ± | 8.46E-01 | 1.75E-01 | ± | 1.50E-02 | 2.95E-01 | ± | 1.95E-02 | | | |
| Cu ion | | | | | | | | | | | | | | | | |
| Energy | Charge state | Zn K [b/sr] | | | Ti K [b/sr] | | | Fe K [b/sr] | | | Ta L [b/sr] | | | | | |
| 40 | 7 | 1.79E+02 | ± | 1.22E+01 | 1.06E+03 | ± | 6.21E+01 | 6.97E+02 | ± | 7.87E+01 | 4.27E+02 | ± | 5.39E+01 | | | |
| 35 | 6 | 1.28E+02 | ± | 8.69E+00 | 8.39E+02 | ± | 4.78E+01 | 5.54E+02 | ± | 6.65E+01 | 3.18E+02 | ± | 3.29E+01 | | | |
| 30 | 5 | 9.30E+01 | ± | 6.34E+00 | 6.58E+02 | ± | 3.84E+01 | 4.52E+02 | ± | 4.85E+01 | 2.02E+02 | ± | 3.37E+01 | | | |
| 25 | 4 | 5.73E+01 | ± | 3.90E+00 | 4.80E+02 | ± | 2.80E+01 | 3.33E+02 | ± | 3.65E+01 | 1.21E+02 | ± | 2.45E+01 | | | |
| 20 | 4 | 4.05E+01 | ± | 2.76E+00 | 3.33E+02 | ± | 1.94E+01 | 2.25E+02 | ± | 2.59E+01 | 6.20E+01 | ± | 1.72E+01 | | | |
| 15 | 3 | 2.10E+01 | ± | 1.52E+00 | 1.85E+02 | ± | 1.36E+01 | 1.51E+02 | ± | 2.43E+01 | 2.57E+01 | ± | 1.02E+01 | | | |
| 10 | 3 | 1.08E+01 | ± | 6.79E-01 | 9.39E+01 | ± | 4.79E+00 | 3.93E+01 | ± | 7.49E+00 | 1.23E+01 | ± | 4.18E+00 | | | |
| Br ion | | | | | | | | | | | | | | | | |
| Energy | Charge state | Zn K [b/sr] | | | Ti K [b/sr] | | | Fe K [b/sr] | | | Ta L [b/sr] | | | | | |
| 40 | 7 | 1.72E+02 | ± | 1.13E+01 | 2.57E+02 | ± | 1.61E+01 | 2.93E+02 | ± | 3.03E+01 | 2.32E+02 | ± | 1.72E+01 | | | |
| 35 | 7 | 1.39E+02 | ± | 5.99E+00 | 2.23E+02 | ± | 1.20E+01 | 2.57E+02 | ± | 2.34E+01 | 2.04E+02 | ± | 1.34E+01 | | | |
| 30 | 7 | 1.13E+02 | ± | 4.35E+00 | 1.67E+02 | ± | 1.73E+01 | 1.97E+02 | ± | 1.35E+01 | 1.41E+02 | ± | 1.83E+01 | | | |
| 25 | 5 | 5.22E+01 | ± | 3.03E+00 | 8.01E+01 | ± | 5.17E+00 | 8.99E+01 | ± | 9.58E+00 | 6.33E+01 | ± | 4.51E+00 | | | |
| 20 | 5 | 4.09E+01 | ± | 1.84E+00 | 4.67E+01 | ± | 2.47E+00 | 5.76E+01 | ± | 3.61E+00 | 3.73E+01 | ± | 2.54E+00 | | | |
| 15 | 3 | 2.02E+01 | ± | 8.52E-01 | 2.13E+01 | ± | 1.09E+00 | 2.32E+01 | ± | 1.46E+00 | 2.07E+01 | ± | 1.36E+00 | | | |
| 10 | 3 | 9.56E+00 | ± | 3.66E-01 | 0.00E+00 | ± | 0.00E+00 | 1.02E+01 | ± | 6.29E-01 | 1.30E+01 | ± | 8.42E-01 | | | |

**Table II.** Differential cross sections for the ion-target combinations detailed in the text.

FIGURES

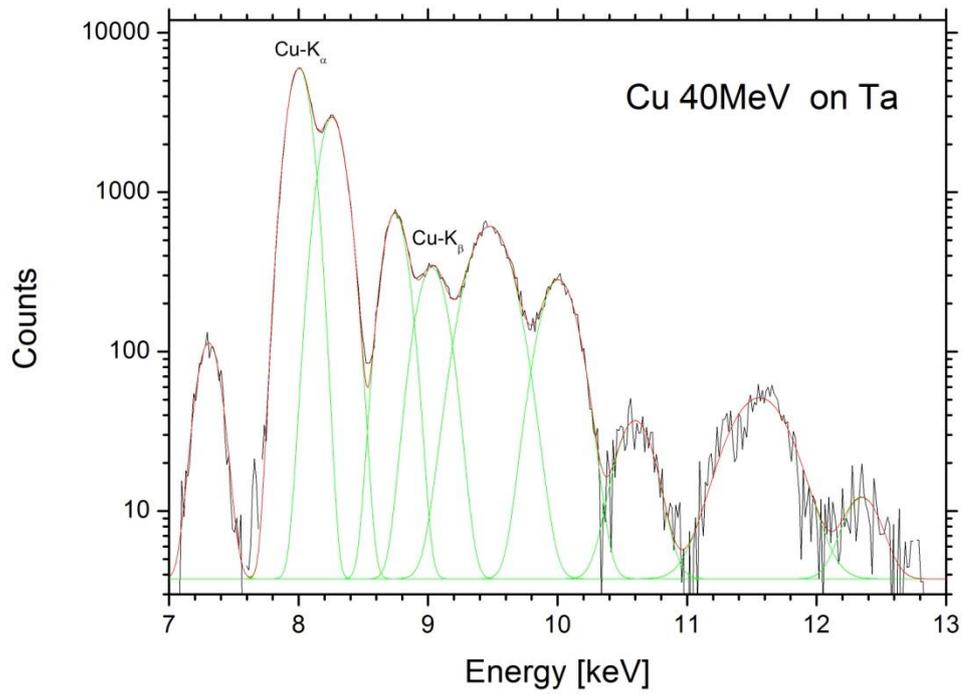

**Figure 1.** Example of X-ray spectrum deconvolution. The case of 40 MeV Cu ions on a Ta target is shown.

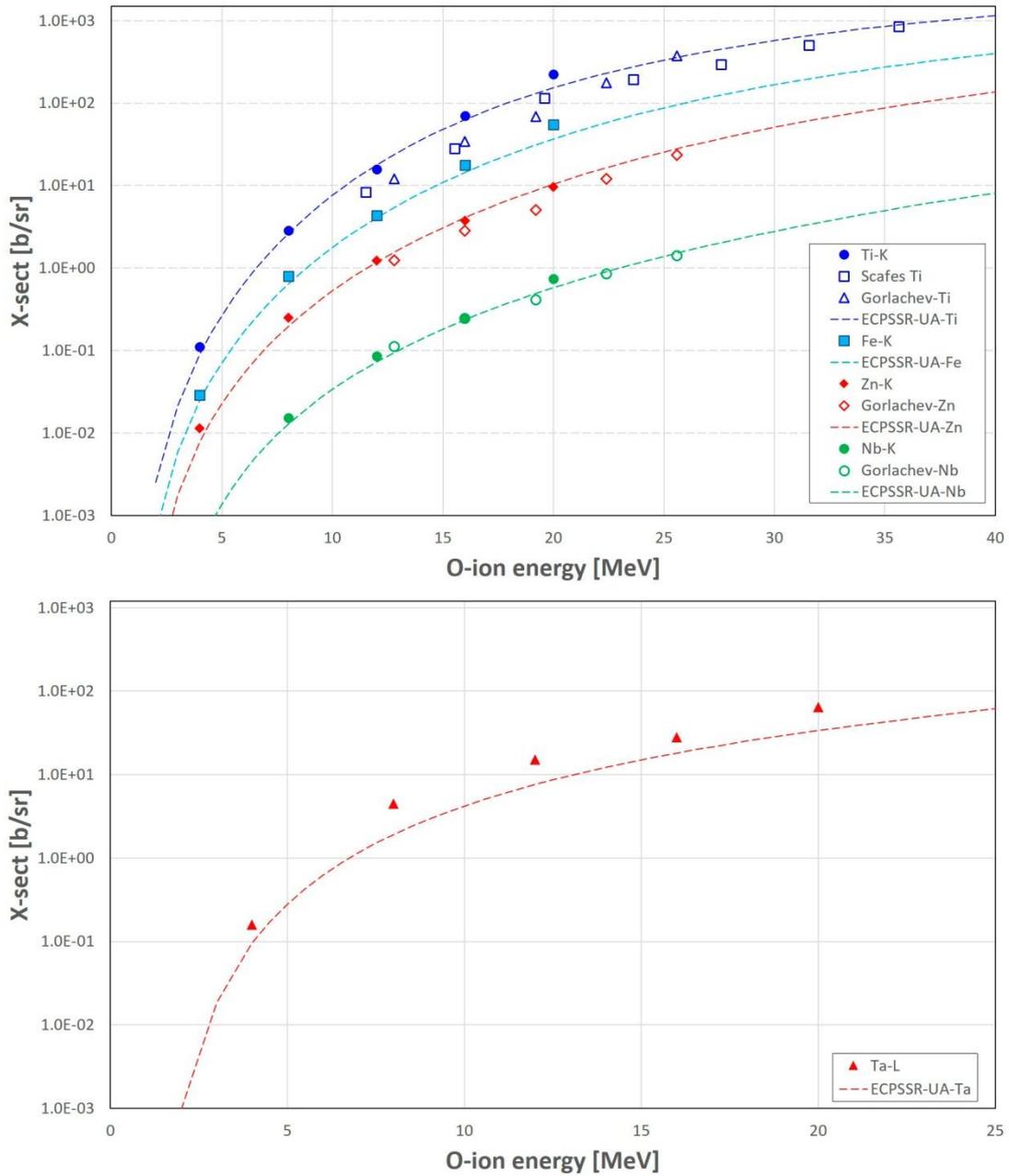

**Figure 2.** Top: X-ray production differential cross section for the case of oxygen ions and the K-lines of Ti, Fe, Zn and Nb. Bottom: the same for the L-lines of Ta. Data published by other authors have been included for comparison. Error bars are not shown for simplicity. ECPSSR-UA calculations are compared to the data.

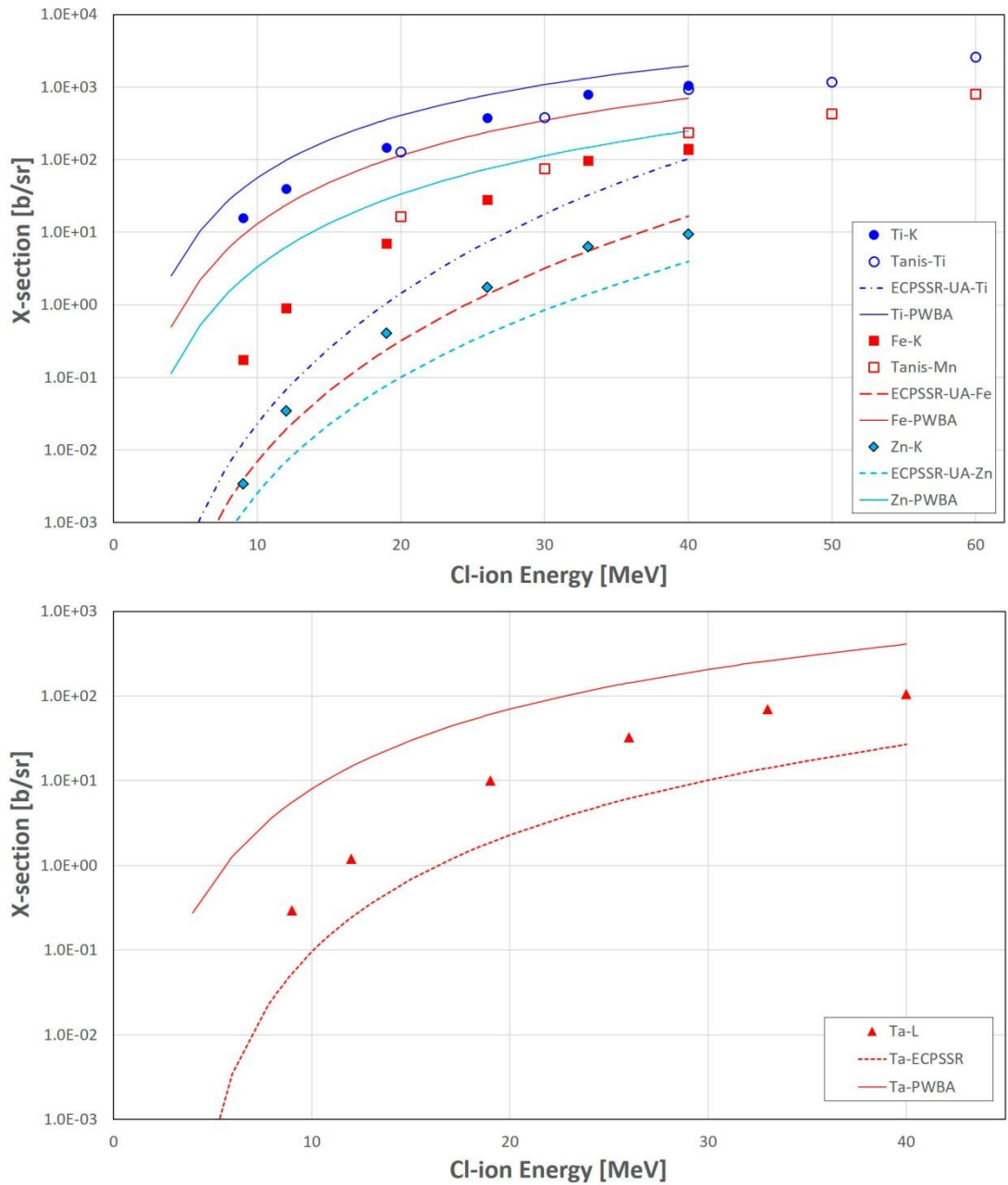

**Figure 3.** Top: X-ray production differential cross section for the case of chlorine ions and the K-lines of Ti, Fe, and Zn. Bottom: the same for the L-lines of Ta. Data published by other authors have been included for comparison. Error bars are not shown for simplicity. ECPSSR-UA (dashed lines) and PWBA calculations (full lines) are compared to the data.

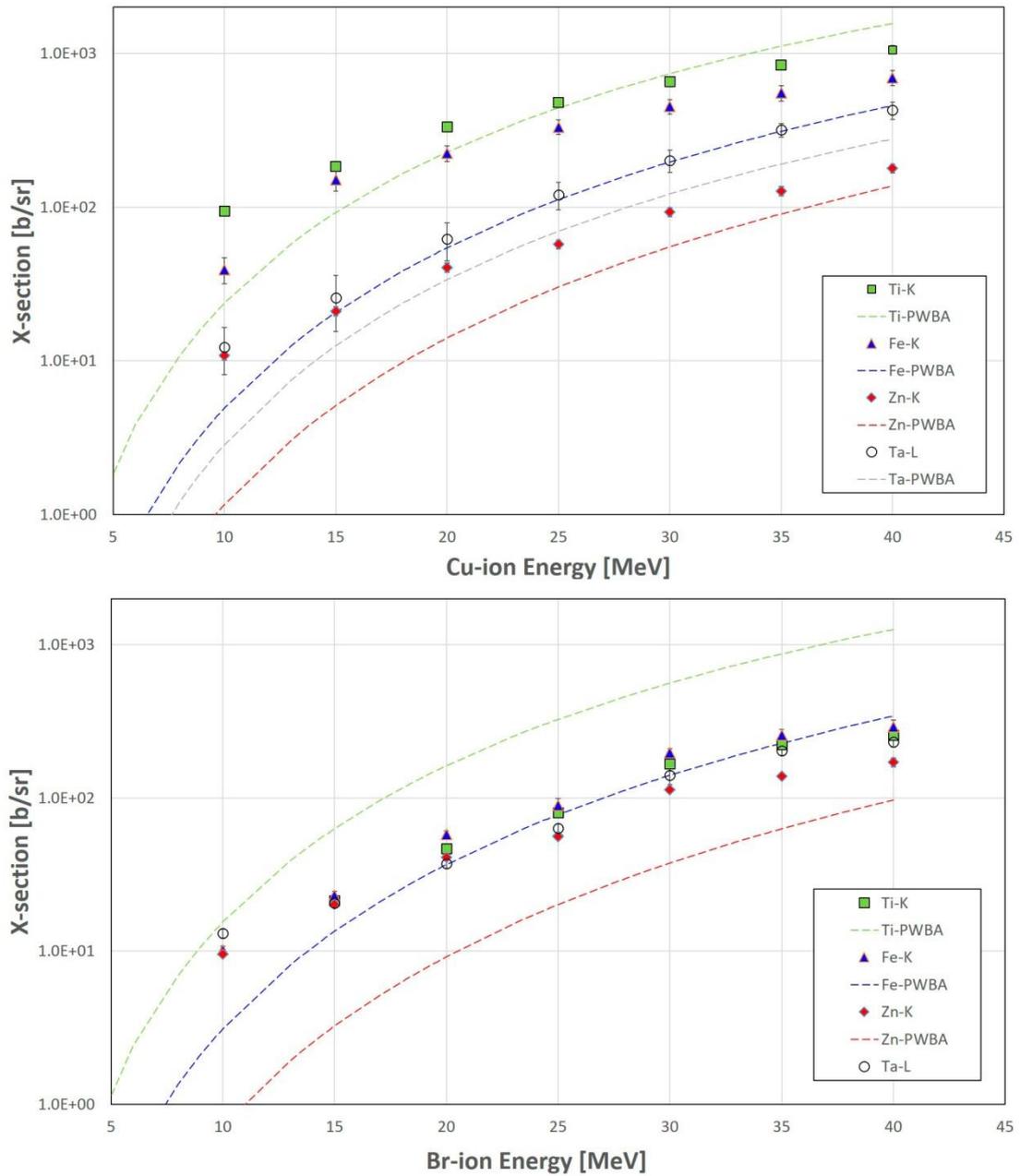

**Figure 4.** Top: X-ray production differential cross section for the case of copper ions and the K-lines of Ti, Fe, and Zn as well as the L-lines of Ta. Bottom: the same for the case of Br ions. Error bars are not shown for simplicity. PWBA calculations are compared to the data.